\documentclass[aps,prl,twocolumn,superscriptaddress,showpacs]{revtex4}
\usepackage{graphicx}

\begin{document}

\title{Long lived acoustic vibrational modes of an
embedded nanoparticle}

\author{Lucien Saviot}
\affiliation{Laboratoire de Recherche sur la R\'eactivit\'e des Solides,
UMR 5613 CNRS - Universit\'e de Bourgogne\\
9 avenue A. Savary, BP 47870 - 21078 Dijon - France}
\email{lucien.saviot@u-bourgogne.fr}

\author{Daniel B. Murray}
\affiliation{Department of Physics and Astronomy,
Okanagan University College, 3333 College Way,
Kelowna, British Columbia, Canada V1V 1V7}
\email{dmurray@ouc.bc.ca}

\date{\today}

\begin{abstract}
Classical continuum elastic calculations show that the
acoustic vibrational modes of an embedded nanoparticle can
be lightly damped even when the longitudinal plane wave
acoustic impedances $Z_o=\rho v_L$ of the nanoparticle and
the matrix are the same.  It is not necessary for the matrix
to be less dense or softer than the nanoparticle in order to
have long lived vibrational modes.  A corrected formula for
acoustic impedance is provided for the case of longitudinal
spherical waves.  Continuum boundary conditions do not
always accurately reflect the microscropic nature of the
interface between nanoparticle and matrix, and a multi-layer
model of the interface reveals the possibility of additional
reduction of mode damping.
\end{abstract}

\maketitle

   A classical vibrational mode weakly coupled to a cold environment
will have an amplitude exponentially decaying with time. The vibrational
frequency of the mode may be slightly shifted from what it would have
been if the environmental coupling is reduced to zero. When coupling
is stronger, the mode loses its individual identity, and it becomes
necessary instead to consider normal modes of the entire system.

Nearly spherical clusters of atoms a few nanometers in
size (nanoparticles) have mechanical vibrations which can be
initiated through sudden thermal excitation with a brief laser
pulse.\cite{VoisinJPCB01} Subsequent evolution of the vibration can be
followed by observing the amount of light absorption by a probe pulse
following the pump pulse by a controlled time.

It is rarely possible to perform such experiments on
isolated nanoparticles.  Normally there is mechanical
coupling to an environment.  This can be (1) a powder of
nanoparticles (where adsorption is important) (2) a
substrate to which the nanoparticle is attached (3) liquid
solution or (4) a solid surrounding matrix, usually glass.
Propagation of acoustic waves away from the nanoparticle is
often the dominant mechanism for mechanical energy loss.  As
a result, vibrational modes are damped.

The organization of this paper is as follows:  We begin with
the current ``state of the art'' of vibration eigenmode
calculation for free and matrix-embedded spheres. Then we
focus on a particular material combination which gives rise
to very low damping of the fundamental ``breathing'' mode
even though the longitudinal plane wave acoustic impedances
in the two materials are the same. We then derive the
correct expression for the acoustic impedance for such modes
and explain this result.  Finally, we extend the calculation
to more realistic materials where the sphere-matrix contact
is not perfect and to modes having different symetries.

A quantum mechanical analysis of energy loss of a
nanoparticle weakly coupled to a substrate heat bath has
been done.\cite{PattonPRB03}  Otherwise, the nanoparticle
has been approximated as a classical isotropic homogeneous
linear elastic continuum
object in contact with a similarly idealized zero temperature matrix.
We consider such a spherical nanoparticle of
radius $R_p$, density $\rho_p$ and speeds of sound $v_{Lp}$ and
$v_{Tp}$ embedded within a macroscopic matrix of density
$\rho_m$ and speeds of sound $v_{Lm}$ and $v_{Tm}$.
Idealized continuum boundary conditions\cite{Bettenhausen03}
of continuous displacement and normal stress components have
been applied where the nanoparticle contacts the matrix,
but these will be reaxamined later on.

``Pseudomodes'' of a nanoparticle embedded in a solid or
fluid matrix can be found\cite{dubrovskiy81} yielding
complex valued frequencies $\omega^{\mathrm{CFM}}_{q \ell n }$
of the ``complex frequency model'' (CFM).
$q$ can be torsional (TOR) or spheroidal (SPH).
$\ell$ is the angular momentum and $m$ is the $z$-component.
$n \geq 0$ is the mode index.
$Re(\omega^{\mathrm{CFM}}_{q \ell n }) = 2 \pi / T$ are
shifted from $\omega^{\mathrm{FSM}}_{q \ell n }$ of a free sphere
model (FSM) as first calculated in 1882 by
Lamb\cite{lamb1882} and some new modes
appear.\cite{murrayPRB03}
$Im(\omega^{\mathrm{CFM}}_{q \ell n }) = 1 / \tau = \frac12 \Delta \omega$
where $\tau$ is the damping time and $\Delta \omega$ is the
full width at half maximum (FWHM) of a Brillouin or Raman peak.
The quality factor of each pseudomode is defined as
$Q=\omega / \Delta \omega$.
This paper is mainly focused on (SPH,$\ell = 0$) modes which are the
only ones to have been experimentally observed in pump-probe
experiments. These modes are also Raman active.

Interpretation of CFM is not
straightforward, since the resulting pseudomodes are not
orthonormalizable and blow up exponentially with the radial
coordinate.  This makes a correspondence to quantum theory
unclear.  An alternative is to calculate all of the normal
modes of a nanoparticle (``core'') surrounded by a large
spherical matrix (``shell''), imposing an arbitrary
condition on the matrix outer surface.\cite{PortalesPRB02}
The resulting core-shell model (CSM) is conservative and
has real valued mode frequencies.\cite{murrayPRB03}
However, there is a continuum of normal modes in the
macroscopic limit that the matrix radius is infinite.  Note
that highly damped ``matrix modes''\cite{murrayPRB03} are
not considered in this work.

The connection between the continuum of CSM modes and the
discrete spectra of CFM pseudomodes is clearly seen by
making a plot of the mean squared displacement within the
nanoparticle $\langle u^2 \rangle_p$ as a function of
frequency for CSM modes.\cite{murrayPRB03}  The resulting
peaks correspond closely to $Re(\omega^{\mathrm{CFM}}_{q \ell n })$
and the FWHM of the CSM peaks corresponds closely to
$2 Im(\omega^{\mathrm{CFM}}_{q \ell n })$.

Even without applying the complexities of CFM or CSM, the
qualitative effect of the matrix in shifting and broadening
FSM frequencies can often be correctly anticipated by
comparing the longitudinal plane wave acoustic impedances
(LPWAI) $Z_{op} = \rho_p v_{Lp}$ and
$Z_{om} = \rho_m v_{Lm}$.

The ``rule of thumb'' is that nanoparticle vibrational modes
will be lightly damped when the LPWAI's satisfy
$Z_{op} \gg Z_{om}$ or $Z_{op} \ll Z_{om}$.  Otherwise,
similar LPWAI's of the nanoparticle and matrix allow
vibrational energy to freely flow out, resulting in strong
damping.

Here we report dramatic exceptions to this.  It is possible
for a nanoparticle to have lightly damped 
normal modes without LPWAI's $Z_{om}$ and $Z_{op}$ being very different,
with the extreme possibility of $\tau$ being large
even if $Z_{om} = Z_{op}$.
Furthermore, lightly damped nanoparticle frequencies can
differ greatly from $\omega^{\mathrm{FSM}}_{q \ell n }$.
These possibilities have remained
obscure up to now because experiments were
limited primarily to situations in which the LPWAI's satisfied
$Z_{op} > Z_{om}$ for
which FSM works extremely well.  Furthermore, nanoparticle
size variation within a sample creates extrinsic broadening
of the modes which can sometimes exceed intrinsic broadening.

An example will illustrate when such unexpected
results can occur: a gold nanoparticle embedded
in a diamond matrix.  We use the elastic
constants of bulk fcc gold at 300~K\cite{LBAu} and
use directional averaging\cite{saviotPRB03} to
obtain longitudinal and transverse speeds of sound of
3330~m/s and 1250~m/s respectively.  We use elastic constants
for bulk crystalline diamond at 300~K\cite{McSkimin}
and similarly obtain speeds of sound 18190~m/s and 12350~m/s.
The resulting LPWAI's are identical:
$Z_{op} \approx Z_{om} \approx 6.4 \cdot 10^7 \textrm{kg m}^{-2}\textrm{s}^{-1}$.
In the absence of any apparent impedance mismatch, the ``rule of thumb''
says that the vibrational modes of the nanoparticle would become so
broadened by the presence of the matrix as to be unrecognizable.

\begin{figure}[!ht]
 \includegraphics[width=\columnwidth]{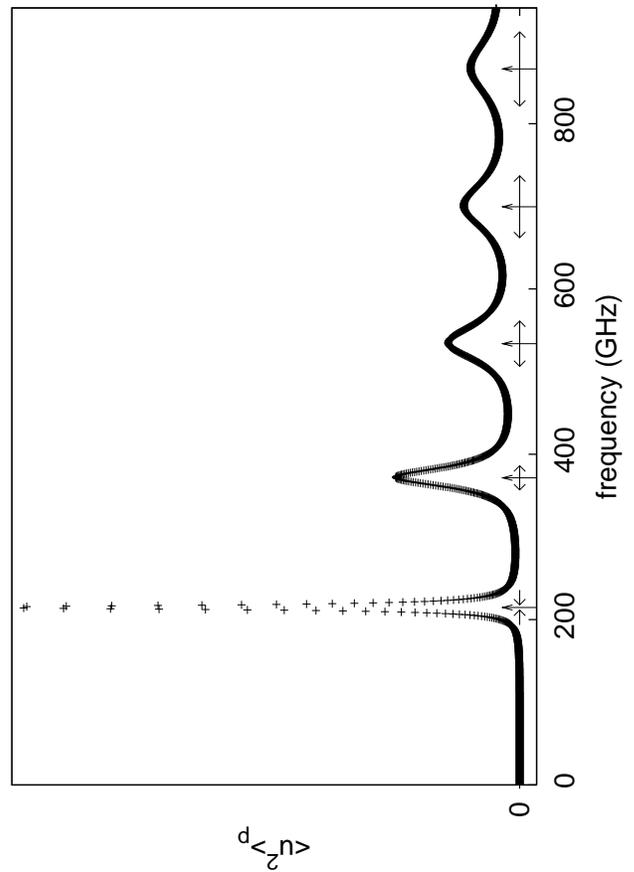}
 \caption{\label{AuDiamond}Mean square displacement within the nanoparticle
 interior of CSM (SPH,$\ell=0$) vibrational modes for Au
 (radius $R_p$=10 nm) in diamond (radius $R_m=2000 \cdot R_p$).
 Arrows indicate positions and FWHM obtained with the CFM.}
\end{figure}

Figure~\ref{AuDiamond} shows something completely different.
$\langle u^2 \rangle_p$ for CSM (SPH,$\ell = 0$)
modes is plotted
versus frequency.  For comparison, $\omega^{\mathrm{CFM}}_{q \ell n }$
are plotted showing $Re(\omega)$ as a vertical arrow with
horizontal arrows to indicate $Im(\omega)$ which
is half the FWHM. The resulting CSM peaks and $Re(\omega^{\mathrm{CFM}}_{q \ell n })$
match very closely.  The gold nanoparticle has several lightly
damped modes of vibration.

\begin{figure}[!ht]
 \includegraphics[width=\columnwidth]{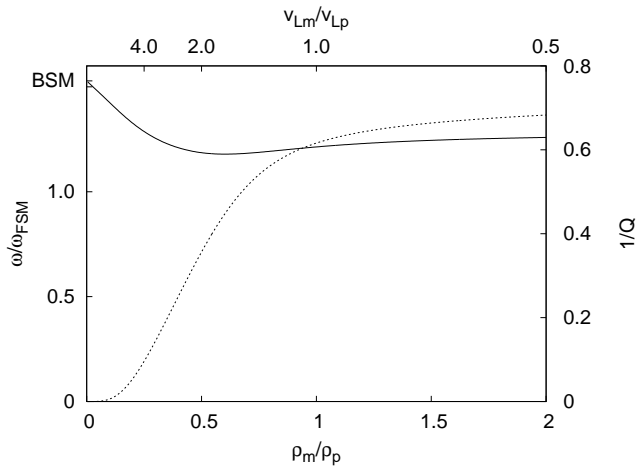}
 \caption{\label{Z1}Normalized CFM frequency (line, left axis)
 and $1/Q=\Delta \omega / \omega$ (dots, right axis)
 of the fundamental (SPH,$\ell = 0$) mode along
 $Z_{om}/Z_{op}=\rho_mv_{Lm}/\rho_pv_{Lp}=1$ where $v_{Tp}/v_{Lp}=0.375$ (gold)
 and $v_{Tm}/v_{Lm}=0.6$. $\Delta \omega$ is the full width at half maximum.}
\end{figure}

Figure~\ref{Z1} shows how misleading naive comparison of LPWAI's can be.
It was obtained by varying both $\rho_m$ and $v_{Lm}$ while keeping
$Z_{om}=Z_{op}$.
Mode broadening is large only when $v_{Lm} < v_{Lp}$.
Note that for a gold nanoparticle in diamond,
$v_{Tm}/v_{Lm} \approx 0.68$ and $\rho_m/\rho_p \approx 0.18$.
For low matrix mass density, the frequency approaches
that of the bound sphere model (BSM), $\omega^{\mathrm{BSM}}_{\mathrm{SPH}00}$,
which is for a sphere with zero displacement boundary conditions.

For a purely longitudinal spherical wave in an isotropic elastic
medium, the acoustic impedance is $Z = -\sigma_{rr} / v_r$ where stress
component $\sigma_{rr} = 2 \rho v_{T}^2 e_{rr} + \rho (v_{L}^2 - 2
v_{T}^2) ( e_{rr} + e_{\theta \theta} + e_{\phi \phi} )$ and velocity
$v_r = \partial u_r / \partial t$. For this case strain $e_{rr} =
\partial u_r / \partial r$ and $e_{\theta \theta} = e_{\phi \phi} = u_r
/ r$. The radial displacement is $u_r = e^{- i \omega t}(d/dr)(e^{i k_L
r}/r)$ where $\omega = v_L k_L$. The resulting expression for impedance
is
\begin{equation}
Z(s) = \frac{ \rho ( v_{L}^2 + 4 v_{T}^2 ( i s - s^2 ) ) }{v_L (1 + i s)}
\label{eqZ}
\end{equation}
where $s = 1/(k_L r)$.
This can only be applied to the case of (SPH,$\ell = 0$) modes.
In the plane wave limit (large $r$) $Z \rightarrow \rho v_L$.
For a fluid medium, $v_T = 0$ so that $Z = \rho v_L / (1 + i s)$,
a result familiar for sound waves\cite{Kinsler}.
Another interesting case is Poisson ratio $\nu = \frac{1}{3}$
(approximately true of many materials)
where $v_T = \frac{1}{2} v_L$ so that $Z = \rho v_L$ to linear
order in $s$.
Note that $v_T/v_L = \sqrt{\frac{ 1 - 2 \nu}{2 - 2 \nu }}$. 

If $s$ is small, then
$Re(Z) \approx \rho v_{L}$ and
$Im(Z) \approx s \rho v_{L} ( 4 ( v_{T} / v_{L})^2 - 1 )$.
Note that if $v_{T} = \frac{1}{2} v_{L}$ then $Im(Z)$ is
approximately zero.

Define $\alpha = v_{Lm} / v_{Lp}$.
Let $Z_p( 1 / \xi )$ and $Z_m( \alpha / \xi )$ denote the acoustic impedance
evaluated at $R_p$ of the nanoparticle and matrix respectively
for outgoing waves of dimensionless frequency $\xi = \omega R_p / v_{Lp}$.
Note that for a given pseudomode, these
impedances do not depend on $R_p$.
If $Z_p( 1 / \xi ) = Z_m( \alpha / \xi )$ then a longitudinal travelling
wave can cross the nanoparticle-matrix interface without any
reflection.

Consider $Z_p( 1 / \xi )$: for the (SPH,$\ell = 0$,$n$) FSM mode,
$s(r=R_p) \sim \frac{1}{(n+1)\pi}$.
For the fundamental ($n = 0$) mode $s(r=R_p)$ is not small and $Z_p( 1 / \xi )$
is complex-valued.  But for higher order modes $s$ becomes small
so that $Z_p( 1 / \xi ) \rightarrow \rho_p v_{Lp}$. Thus, it is
expected that the acoustic impedance mismatch at the
nanoparticle-matrix interface will be different for higher
order modes.
Figure \ref{AuDiamond} confirms this. Also, as
predicted in a previous paper\cite{murrayPRB03}, the area under
each pseudomode is roughly constant in the $\langle u^2 \rangle_p$
plot. This is important because it means that less damped modes
compensate their lower frequency broadening with a higher mean
square displacement and therefore can have similar electron-phonon
coupling inside the nanoparticle as more damped modes.

Consider the longitudinal radial ($r$) velocity within the
nanoparticle ($p$) to be the sum of inward ($i$) and outward ($o$)
travelling waves:
\begin{equation}
v_{rpo}(x) = e^{-i \omega t} A_{po} \omega e^{ix} ( ( 1/x ) + ( i / x^2 ) )
\end{equation}
\begin{equation}
v_{rpi}(y) = e^{-i \omega t} A_{pi} \omega e^{iy} ( ( 1/y ) + ( i / y^2 ) )
\end{equation}
where $x =  \omega r / v_{Lp}$ and $y = - \omega r / v_{Lp}$.
$A_{po}$ and $A_{pi}$ are amplitudes.
To avoid singularity at the origin, $A_{po} = -A_{pi}$.
Matching of boundary conditions at the interface at $R_p$ corresponds
to the equation:
\begin{equation}
\frac{ Z_{p}(1/\xi) - Z_{m}(\alpha/\xi) }{ Z_{p}(-1/\xi) + Z_{m}(\alpha/\xi) }
=
\frac{v_{rpi}(-\xi)}{v_{rpo}(\xi)}
\end{equation}
The roots of this equation are the complex valued
dimensionless frequencies
$\xi^{\mathrm{CFM}}_{\mathrm{SPH}0n}$.  This equation is
equivalent to previous solutions\cite{dubrovskiy81} of CFM
for the (SPH,0) case, but clearly illustrates the role of
the acoustic impedances in determining the mode frequencies.

Previous studies\cite{VoisinPhysicaB02} were restricted
to analyzing the situation of a soft, light matrix where
$v_{Lm} < v_{Lp}$ and $\rho_m < \rho_p$, and found that
$\tau$ does not depend on $n$.  But we find that in other
situations damping can vary dramatically with mode order,
as clearly seen in Fig.~\ref{AuDiamond}. 

In the matrix, $s = \alpha / \xi$ can be small if
$v_{Lm} \ll v_{Lp}$.  But if $v_{Lm} \gg v_{Lp}$ then $s$
can be large, and in this limit
$Z_m(\alpha / \xi ) \rightarrow 4 i \rho_m v_{Tm}^2 / ( \omega R_p )$,
and the effect of the matrix is a purely restorative force,
as if the matrix is a massless spring.  Energy cannot be
absorbed by the matrix in this limit, so that
$\tau \rightarrow \infty$ when $v_{Lp} / v_{Lm} \rightarrow 0$.

We have explored the lowest (SPH,$\ell = 0$) mode (ignoring
highly damped ``matrix modes'' when they are present) in the
parameter space of $\rho_p$, $\rho_m$, $v_{Lp}$, $v_{Tp}$,
$v_{Lm}$ and $v_{Tm}$ and found some general patterns.
Notably, good impedance matching between nanoparticle and
matrix as exhibited in very strong lowest (SPH,0) mode
damping is an unusual occurence requiring special conditions.
Closeness of the LPWAI's alone is not sufficient.

If $\nu_m \ll \nu_p$ (where $\nu_m$ and $\nu_p$ are the
Poisson ratios of the matrix and particle respectively) then
there is never strong damping, but we can have strong
damping if $\nu_{m} \gg \nu_{p}$ such as for a liquid matrix.
Assuming the Poisson ratios are close, it is usually
necessary that $\rho_m \geq \rho_p$ and $v_{Lm} \leq v_{Lp}$.
Under these conditions, $s$ is small in the matrix and the
LPWAI's are close to the true impedances.  In addition, if
$\nu_m$ and $\nu_p$ are both close to $\frac13$ then strong
damping is still possible if $\rho_m$ is as little as
$0.8 \rho_p$.  Conversely, light damping is always seen when
any of the following conditions is satisfied:
(1) $\nu_p \gg \nu_m$
(2) $\rho_m \ll 0.8 \rho_p$
(3) $v_{Lm} \gg 1.2 v_{Lp}$.

As was shown in a previous work,\cite{PortalesPRB02} the
microscopic details of the interface (\textit{e.g.} the
specific nature of the mechanical connection of the gold
crystal lattice to the diamond matrix) can tend to isolate
(\textit{i.e.} decouple) the nanoparticle from the
surrounding medium.  In particular, continuum-type boundary
conditions applied to the interface may not be
optimal.\cite{Bettenhausen03}  One rough-and-ready way to
adapt the CFM and CSM models to tentatively take this
into account is to introduce an intermediate shell layer
(X-layer) of inner radius $R_p - d_X / 2$, outer radius
$R_p + d_X / 2$,
density $\rho_X$ and speeds of sound $v_{LX}$ and $v_{TX}$.
Depending on the elaboration process, the nanoparticle and
the surrounding matrix do not necessarily tightly adhere.
Different expansion coefficients for example could result in
a void in the matrix bigger than the nanoparticle and the
nanoparticle being only partially in contact with the matrix.
The X-layer is the most elementary way to explore the effect
of deviations from continuum boundary conditions.  Such
calculations are relatively easy to perform using CFM or CSM.

It is important to demonstrate the robustness of CFM results
against the microscopic details of the interface.  The
nature of the interface is unknown, but it can be guessed to
be of lower density, poorly-ordered and not strongly-bonded.
It is plausible to assume a low density and low elastic
constants comparable to water or plastic.  If $d_X$ is thin,
neither $\rho_X$ nor $v_{TX}$ have a strong effect.  What is
important is the elastic constant of the X-layer in terms of
$\rho_X v_{LX}^2$ relative to the thickness of the X-layer,
$d_X$.  It is convenient to introduce the dimensionless
parameter $\sigma_X = d_X \rho_p v_{Lp}^2 / (R_p \rho_X v_{LX}^2)$.
For examples we looked at, the larger $\sigma_X$ is, the
larger $\tau$ and $Q$ become.

A possible example of this may have been seen for gold
nanoparticles embedded in TiO$_2$. This is unrelated
to our previous example of Au in diamond.
Experimental observations of nanoparticle ``breathing modes''
(those with purely radial vibration) for Au nanoparticles in
a TiO$_2$ matrix using the femtosecond pump probe
technique\cite{Qian} reveal surprisingly light damping. Qian
\textit{et al.} obtained an oscillation frequency
of 0.15 THz (5~cm$^{-1}$) with a damping time of 55~ps
($Q \approx 26$) for $R_{p}=10$~nm.  We use TiO$_2$ sound
speeds of 8610~m/s and 5160~m/s and density
4.097~g/cm$^3$.\cite{Minnear77}  Our XCFM calculations for
very small $\sigma_X$ predict
$\tau \approx (10.7 + 21.0 \sigma_X)$~ps
or $Q \approx 5.92 + 11.7 \sigma_X$.
This $\tau$ provides only an upper bound on the experimental
damping time because of dephasing due to particle size
variation.  The large $\tau$ of Au in TiO$_2$ could be
obtained by a water-like layer
between the nanoparticle and the matrix on the order of
0.1~nm thick.  Thus, details of the nanoparticle-matrix
interface at the atomic level are crucial in determining
the damping of embedded nanoparticle vibrational modes.
However, proper first-principles quantitative modelling of
the interface requires consideration of detailed features 
of crystal and electronic properties at the atomic level,
in particular the nature of the bonding.\cite{Carrier02}
For the systems we consider, there is insufficient
information about the interface to attempt such a
calculation, especially since the matrix is amorphous.

The ability to choose the matrix so as to have nanoparticles
with lightly damped vibrational modes is important for
several reasons.  First, by reducing homogeneous broadening,
it allows inhomogeneous broadening effects to stand out
clearly.  This results in a clearer picture for femtosecond
pump-probe experiments where the damping time is longer but
also in low frequency Raman experiments where a better
separation of scattering lines is possible.  Second, as was
shown by the pump-pump-probe experiments in Del Fatti
\textit{et al.}\cite{DelFattiJPCA00} it is possible to
control the acoustic motion of silver nanoparticles.  With
higher $Q$ factor, the first pump pulse would ``launch'' the
motion of every nanoparticles but the next pump pulse could
selectively enhanced the vibration amplitude of a smaller
subset of the nanoparticles.
                         
Solid matrix-embedded nanoparticles always have $Q$ factor lower than
free ones.  However, free nanoparticles suffer from many drawbacks. For
example, surface adsorption can affect their properties.  Also it is
often technologically easier to deal with matrix embedded nanoparticles.
Where solid embedding is technologically desirable, our results provide
guidance about how the matrix can be optimally chosen.

Finally, the sphere radius is irrelevant in this paper ($Q$ does
not depend on $R$).  So there are some potential applications for other
kinds of systems.  Such damped vibrational modes of a virus in water may
be relevant to its mechanical response to ultrasound.\cite{saviotPRE04}

All of the foregoing discussion applies only to spheroidal vibrations
with $\ell = 0$. A similar analysis can be carried out for torsional
modes, but a separate expression for acoustic impedance is required for
each value of $\ell$.  It is not apparent how to make the extension to
spheroidal modes with $\ell \neq 0$.  Calculations made on CdS spheres
embedded in silica in a previous work\cite{murrayPRB03} revealed lightly
damped fundamental torsional modes which we could not understand at
that time.  The situation is even more dramatic for gold spheres inside
diamond for which the lowest torsional modes have very high quality
factors ($Q \sim 900$ for (TOR,$\ell = 1$,$n = 0$), $Q \sim 20000$ for
(TOR,$\ell = 2$,$n = 0$), \ldots)

\bibliography{audi}

\end{document}